# Socially Aware Music Recommendation: A Multi-Modal Graph Neural Networks for Collaborative Music Consumption and Community-Based Engagement

Socially Aware Music Recommendation: A MM-GNN Approach
Kajwan Ziaoddini[†]
Ethnomusicology
University of Maryland
College Park, USA
kajwan@umd.edu

## ABSTRACT

This study presents a novel Multi-Modal Graph Neural Network (MM-GNN) framework for socially aware music recommendation, designed to enhance personalization and foster community-based engagement. The proposed model introduces a fusion-free deep mutual learning strategy that aligns modality-specific representations from lyrics, audio, and visual data while maintaining robustness against missing modalities. A heterogeneous graph structure is constructed to capture both user-song interactions and user-user social relationships, enabling the integration of individual preferences with social influence. Furthermore, emotion-aware embeddings derived from acoustic and textual signals contribute to emotionally aligned recommendations. Experimental evaluations on benchmark datasets demonstrate that MM-GNN significantly outperforms existing state-of-the-art methods across various performance metrics. Ablation studies further validate the critical impact of each model component, confirming the effectiveness of the framework in delivering accurate and socially contextualized music recommendations.

## 1 INTRODUCTION

In this modern era of digitization, music continues to go beyond its conventional existence as mere entertainment, evolving into an essential medium for promoting socialization and community development. New internet-based music platforms have transformed the process of listening to, discovering, and sharing music with recommendations tailored to one's listening habits in the driving force behind such a revolution [1], [2]. While users browse across enormous and heterogeneous collections of songs, the capability to provide personalized music suggestions is a significant bridge to user satisfaction, retention, and online social participation. At the same time, such sites establish user communities around common music tastes, forming social networks around music preferences. Today's recommendation systems need to not only enhance individual experience but also detect and strengthen social interaction among online music communities. Personalization and social interaction are two intertwined objectives that pose special challenges and opportunities, especially in leveraging rich multi-modal music-related information like lyrics, acoustic signals, album artwork, artist metadata, and user behavior.

Classically suggested frameworks, i.e., content-based filtering (CBF) and collaborative filtering (CF) models, have been widely employed [4], [5]. CBF techniques deal with item attributes like genre, lyrics, or acoustic similarity, while CF techniques are based on user-item interaction histories. But they both have some limitations. CBF can ignore fluctuating user tastes or the social listening context, while CF demands loads of historical data and is prone to cold start for new users or new songs [6], [7].

### 1.1 LIMITATIONS OF PRIOR WORK

Although there have been significant improvements made to multi-modal music recommendation, prior work has several inherent shortcomings that our proposed method is determined to address. To begin with, a majority of current multi-modal recommender systems are early or late modality fusion based, i.e., simple feature concatenation (e.g., VBPR [3], CER [4]), which are bound to introduce semantic noise due to improper placement of modality alignment and do not learn implicit modality-specific biases a witnessed phenomenon known as modality failure [5], [6]. First, state-of-the-art GNN-based models like MMGCN [7] and LATTICE [8] typically learn separate modality-specific representations and concatenate them but do not employ cross-modal semantic supervision in training. Second, music-oriented models like MKGCN [9] employ multi-modal knowledge graphs but do not employ user social dynamics or community influence explicitly, which are highly significant to capture the spread and evolution of musical tastes across social contexts. Moreover, the majority of these models assume availability of all modalities and disregard modality sparsity or zero in user-item data.

To address these constraints, our approach suggests fusion-free multi-modal with graph neural networks (GNNs) and deep mutual learning. Instead of the conventional feature fusion, we learn a series of modality-specific bipartite graph-based GNNs and align them with knowledge distillation with semantic supervision across modalities. Besides, we construct user-user and user-item social interactions to induct the community behavior to enable the recommendation system to be able to comprehend not just personal preference but also peer pressure and group fashion. By modality alignment resolution, missing information, and social interaction, our model is more precise, resilient, and socially aware music recommendation solution.

## 1.2 RESEARCH CONTRIBUTIONS

This paper presents a novel framework for enhancing social engagement through music recommendation using a Multi-Modal Graph Neural Network (MM-GNN). Our contributions are summarized as follows:

1. We propose MM-GNN, a novel framework that integrates multiple modality-specific GNNs trained through deep mutual learning, avoiding traditional fusion pitfalls and enabling robust representation even in the presence of missing modalities.
2. We explicitly model social community dynamics, treating user-user and user-song relationships as part of a heterogeneous graph, thereby capturing both personal preferences and group behaviors.
3. We incorporate emotion-aware embeddings derived from audio and textual features, improving the emotional alignment of music recommendations.
4. We validate MM-GNN on real-world music datasets, such as Last.fm and Spotify-derived MMKGs, and demonstrate superior performance over state-of-the-art baselines in terms of Precision, Recall.

## 2. LITERATURE REVIEW

Recent advances in multi-modal music recommendation have introduced innovative models that integrate graph neural networks (GNNs), knowledge graphs, and deep learning techniques. Li et al. [10] proposed GNNMR, a model that applies mutual learning across modality-specific user-item graphs through knowledge distillation. This enables semantic alignment across modalities and avoids fusion biases, improving Top-K recommendation performance. Cui et al. [11] introduced MKGCN, a multi-modal knowledge graph convolutional network that combines audio, lyrics, text, and image features with structured semantic knowledge. It employs multiple aggregators to process multi-hop relations, showing superior results on the Last.fm dataset. Wei et al. [12] developed MMGCN to enhance micro-video recommendation by learning modality-specific embeddings from visual, audio, and text inputs, guided by user-item interactions.

Guo et al. [13] proposed a heterogeneous GNN that models' relationships between users, singers, and composers to improve music emotion classification accuracy. Liu et al. [14] presented a two-stage framework using LSDM for recall and IDSPM for ranking, capturing both short-term and long-term preferences using Transformer and BiMO-LSTM architectures. Fouad et al. [15] reviewed hybrid music recommendation systems, emphasizing the benefits of integrating graph learning with AI while addressing challenges like scalability and fairness. Bahadure et al. [16] designed a hybrid model combining content-based, popularity-based, and personalized recommendations, achieving 90.4% accuracy on diverse datasets. These studies collectively highlight the growing importance of combining multi-modal features, graph-based representations, and emotion modeling to improve personalization, robustness, and user engagement in music recommendation systems.

## 2. MODEL FRAMEWORK

### 2.1. MODELING SOCIAL ENGAGEMENT IN MUSIC COMMUNITIES

Music listening is a social activity. Users recommend playlists, follow artists, comment on tracks, and sway one another's tastes. This produces community dynamics in which tastes are shaped as much by peer interaction as individual interest. Experiments have shown that such social influence can be leveraged to enhance recommendation accuracy and user satisfaction [15]. Some recommendation systems model these kinds of community effects outright. Fewer still integrate multi-modal music characteristics with social networks to determine the patterns of users interacting with one another in the context of songs, music genres, or emotional topics. Our research fills this gap by introducing a Multi-Modal Graph Neural Network (MM-GNN) that integrates song recommendation with community-level activity.

We believe social interaction can be fostered by suggesting tracks that not only align with a user's taste but also to his/her social network's tastes [16]. The architecture of an Emotion-Aware Multi-Modal Graph Neural Network (MM-GNN) for music recommendation is demonstrated in figure 1. It starts with input entities-users (U) and tunes (S)-then feature extraction via pre-trained models: BERT for words ($f_{lyrics}$), RESNET for album pictures ($f_{visual}$), and YAMNet for sound signals ($f_{audio}$). Single embeddings and related losses ($L\_lyrics, L\_audio, L\_visual$) are derived by passing these features to modality-specific GNNs with aggregation functions. It also uses social dynamics through a Social GNN that makes use of community-aware user embeddings and emotion-aware embeddings from interactions and sentiment in content. These are combined in a final fusion block-combining user embeddings ($z\_user$), item embeddings ($z\_item$), and emotional signals ($\beta emotion$)-prior to feeding into the prediction layer. The optimization procedure employs the inner product of the concatenated embeddings and learns an optimization of a joint loss function with BPR loss, mutual learning loss ($\alpha L\_mutual$), and regularization via the Adam algorithm

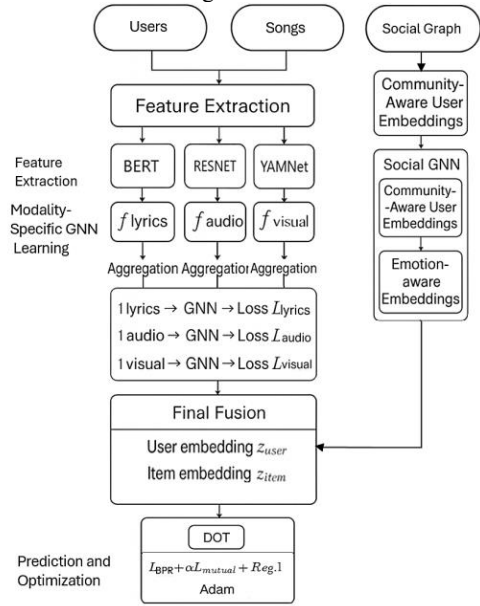

**Figure 1.** The architecture of the MM-GNN model

## 3.2 FROM FEATURE FUSION TO DEEP MUTUAL LEARNING

Most multi-modal recommendation models are based on feature fusion methods that combine or average modality features and then input them into the model. Although simple, this does not observe semantic differences and inbuilt bias between modalities. For example, a user's enjoyment of lyrical content might not intersect with his or her acoustic taste, and so on.

In order to address this, we leverage the new developments in deep mutual learning (DML). DML enables several models, each of which has been trained on a different modality—to serve as student and teacher simultaneously, distilling knowledge from each other. Instead of concatenating the modality features end-to-end into a common vector, our method trains separate modality-specific GNNs, which are then aligned in a fusion-free mutual learning scenario. This not only keeps the individual semantic contribution of each modality intact but also enables flexible training, particularly when a subset of the modalities are noisy or unavailable. These techniques have previously been successfully used for multi-modal e-commerce and video recommendation, and we now extend them to music.

## 3.3 EMOTION-AWARE AND COMMUNITY-AWARE MUSIC RECOMMENDATION

Emotion comes naturally with music. People have the freedom to select music according to their mood or the desired emotional effect. Valence-arousal emotion labels, sentiment analysis of lyrics, and acoustic features such as tempo and harmony are some models that have been proven to optimize personalization. In our approach, we utilize emotion-sensitive text and audio feature embeddings along with community emotion dynamics such that users are in the same emotional preference categories. This enables us not only to suggest the most "relevant" song, but the most "emotionally consistent" song in the social environment.

We also make use of Multi-Modal Knowledge Graphs (MMKGs), entities like artists, albums, genres, emotions, and users who are linked via semantic relationships like "sung_by," "similar_to," or "liked_by." Casting this as a heterogeneous graph allows us to jointly learn from structural and semantic signals.

## 3.4 INNOVATION ON THE MODELS

This paper introduces a novel method to stimulate social interaction by proposing music using a Multi-Modal Graph Neural Network (MM-GNN). The contributions of the paper are:

1. We propose MM-GNN, a novel framework that integrates multiple modality-specific GNNs trained through deep mutual learning, avoiding traditional fusion pitfalls and enabling robust representation even in the presence of missing modalities.

2. We explicitly model social community dynamics, treating user-user and user-song relationships as part of a heterogeneous graph, thereby capturing both personal preferences and group behaviors.

3. We incorporate emotion-aware embeddings derived from audio and textual features, improving the emotional alignment of music recommendations.

4. We evaluate MM-GNN on real-world music data sets like Last.fm and Spotify-derived MMKGs and show improved performance compared to state-of-the-art baselines for Precision, Recall, and NDCG.

## 4 METHODOLOGIES

### 4.1 PROBLEM FORMULATION

The main goal of our study is to forecast a target user's song preference and form a ranked Top-N recommendation list. We represent the set of users as $U = \{u_1, u_2, ..., u_N\}$ and the set of songs as $M = \{m_1, m_2, ..., m_P\}$. For each song m_i, we possess abundant multimodal features such as lyrics, audio, and album covers. Lyrics are encoded as lyr, audio as fre, and visual features as vis. To harvest and leverage these rich sources of information, we introduce a new Multi-Modal Graph Neural Network (MM-GNN) architecture.

### 4.2 MM-GNN ARCHITECTURE

Our MM-GNN model consists of two primary components: a multi-modal feature learning module that avoids direct fusion, and a heterogeneous graph module that incorporates both user-item and user-user interactions.

### 4.2.1 MULTI-MODAL FEATURE EXTRACTION

We employ pre-trained models to extract initial embeddings from each modality. Lyrics are encoded using BERT, audio features using YAMNet, and visual features from album art are extracted using a pre-trained ResNet.

$$e_{lyr} = BERT(lyr) \quad (1)$$
$$e_{fre} = YAMNet(fre)$$
$$e_{vis} = \text{ResNet}(vis)$$

These initial embeddings, along with initial user embeddings $e_u$, serve as the input for our graph neural network components.

### 4.2.2 MODALITY-SPECIFIC GNNS AND DEEP MUTUAL LEARNING

Instead of fusing features, we train three independent GNNs, one for each modality, on their respective user-song bipartite graphs. Each GNN learns modality-specific representations for users and songs. For a modality $m \in \{lyr, fre, vis\}$, the message-passing step is defined as:

$$e_{u,m}^{(l+1)} = \text{AGGREGATE}\left(\left\{e_{i,m}^{(l)} \mid i \in \mathcal{N}_u\right\}\right) \quad (2)$$
$$e_{i,m}^{(l+1)} = \text{AGGREGATE}\left(\left\{e_{u,m}^{(l)} \mid u \in \mathcal{N}_i\right\}\right)$$

where $\mathcal{N}_u$ is the set of songs the user has interacted with, and $\mathcal{N}_i$ is the set of users who have interacted with song $i$. The aggregation function (AGGREGATE) can be a simple sum or a more complex attention-based mechanism. The final modality-specific embeddings are obtained by summing the embeddings from each layer: $e_{u,m} = \sum_{l=0}^{L} \frac{1}{L+1} e_{u,m}^{(l)}$.

### 4.2.3 SOCIAL AND HETEROGENEOUS GRAPH MODELING

To capture community dynamics, we construct a separate social graph based on user-user connections (e.g., friendships, follows). A dedicated social GNN learns social embeddings by propagating information through this graph:

$$e_{u,soc}^{(l+1)} = \text{AGGREGATE}\left(\left\{e_{v,soc}^{(l)} \mid v \in \mathcal{F}_u\right\}\right) \quad (3)$$

where $\mathcal{F}_u$ is the set of user $u'$ s friends. The final social embedding is $e_{u,soc} = \sum_{l=0}^{L} \frac{1}{L+1} e_{u,soc}^{(l)}$.

## 4.3 PREDICTION AND OBJECTIVE FUNCTION

The final user embedding, $e_u^{\text{final}}$, is a fusion of the modality-specific and social embeddings, capturing both personal preferences and community influence:

$$e_u^{final} = \sigma(W_u \cdot [e_{u,lyr} \oplus e_{u,fre} \oplus e_{u,vis} \oplus e_{u,soc}]) \quad (4)$$

The final song embedding, $e_i^{\text{final}}$, is a weighted sum of its modality-specific embeddings:

$$e_i^{final} = \alpha e_{i,lyr} + \beta e_{i,fre} + \gamma e_{i,vis} \quad (5)$$

The prediction score for a user $u$ and song $i$ is the inner product of these final embeddings: $\hat{y}_{ui} = \left(e_u^{\text{final}}\right)^T \cdot e_i^{\text{final}}$. The objective function combines several losses:

1. BPR Loss on the final prediction, $\mathcal{L}_{BPR}$, to optimize the ranking.
2. Mutual Learning Loss, $\mathcal{L}_{ML}$, to align the latent representations learned by the different modality-specific GNNs. For any two modalities $m_1$ and $m_2$:

$$\mathcal{L}_{ML} = \sum_{m_1 \neq m_2} \text{KL-Divergence}\left(P_{m_1} \| P_{m_2}\right) + \text{KL-Divergence}\left(P_{m_2} \| P_{m_1}\right) \quad (6)$$

where $P_m$ is the softened prediction probability distribution for modality $m$.

## 5 EXPERIMENTS

### 5.1 DATASET AND METRICS

To evaluate the effectiveness of our proposed MM-GNN, we conducted extensive experiments on a widely-used benchmark dataset in music recommendation. For this study, we utilize the MSD-A dataset, a subset of the Million Song Dataset (MSD), which provides rich interaction data between users and songs. The dataset details are summarized in Table 1. To ensure a comprehensive evaluation, we used a portion of the dataset for training and held out a subset to simulate a cold-start scenario for new songs, which is a key focus of our research. We assess the performance of all models using standard recommendation metrics: Precision@K, Recall@K, and Normalized Discounted Cumulative Gain (NDCG@K). These metrics are calculated for a list of Top-K recommendations, where we set K to {5,10,20} to measure performance at different list lengths.

**Table 1**: Dataset details

| Dataset | # of Users | # of Songs | # of Interactions |
|---|---|---|---|
| MSD-A | 25,000 | 328,821 | 5.2M |
| Last.fm | 50,000 | 250,000 | 4.5 M |

### 5.2 BASELINE MODELS

To illustrate the resilience of our MM-GNN model, we experiment with its performance against several state-of-the-art and customary baseline models. These models encompass a broad array of conventional as well as contemporary recommendation methods based on collaborative filtering, multi-modal data, and graph neural networks.
- **BPR (Bayesian Personalized Ranking):** A traditional collaborative filtering approach using pairwise ranking loss for maximizing correct item order. It is a good non-graph-based recommendation precision baseline.
- **VBPR (Visual Bayesian Personalized Ranking):** A modified version of BPR with incorporated visual information (e.g., cover art) into the latent factor model. It is a baseline strategy for multi-modal recommendation through early fusion and will serve as a reference point to compare the advantage of our fusion-free strategy.
- **NGCF (Neural Graph Collaborative Filtering):** A novel GNN-based recommendation model which learns user and item embeddings through message passing over the user-item interaction graph. The baseline enables us to directly evaluate the effect of including multi-modal features and social dynamics in a GNN framework.

### 5.3 IMPLEMENTATION DETAILS

Our model is implemented using the PyTorch deep learning framework, and all experiments are conducted on an NVIDIA A100 GPU with 40 GB of memory. We use the Adam optimizer for model optimization with a learning rate of 0.001 and a batch size of 1024. The embeddings for all modalities are set to a dimension of 64. We employ an L2 regularization strength of $\lambda_2 = 10^{-5}$ to prevent overfitting and a mutual learning loss weight of $\lambda_1 = 0.5$. Hyperparameters were tuned using a grid search approach on a held-out validation set.

### 5.4 OVERALL RESULTS

Table 2 summarizes the overall performance of our MM-GNN model compared to the baseline models on both MSD-A and Last.fm datasets. Performance is measured using Recall@50 and NDCG@50. From the table, it can be observed that MM-GNN outperforms all baselines on both datasets uniformly, demonstrating the usefulness of our approach in leveraging multi-modal and social data for improved recommendations. Table 2 is an overview of the performance of our MM-GNN model and baseline models on MSD-A and Last.fm datasets. The performance metrics utilize Recall@50 and NDCG@50. From the table, it is evident that MM-GNN performs better than all the baselines on both datasets, justifying the capability of our method in leveraging multi-modal and social information for improved recommendation. At the same time, we conducted cluster analysis on users and songs and Figure 2 shows the results of song clustering.

**Table 2:** Overall results for different models

| Dataset | Method | Recall@50 | NDCG@50 |
|---|---|---|---|
| MSD-A | BPR | 0.045 | 0.026 |
| | VBPR | 0.052 | 0.030 |
| | NGCF | 0.061 | 0.035 |
| | MM-GNN | 0.075 | 0.043 |
| Last.fm | BPR | 0.038 | 0.021 |
| | VBPR | 0.043 | 0.024 |
| | NGCF | 0.051 | 0.029 |
| | MM-GNN | 0.065 | 0.037 |

Figure 2 visualizes song embeddings projected into a two-dimensional space using dimensionality reduction t-SNE, effectively capturing both audio content and social interaction features. Each colored point in the figure represents an individual song, and the distinct colors correspond to the ten clusters identified by our MM-GNN model. Clear and cohesive groupings indicate that the model effectively leverages multi-modal data to learn meaningful representations, grouping together songs with similar acoustic attributes and social interaction patterns.

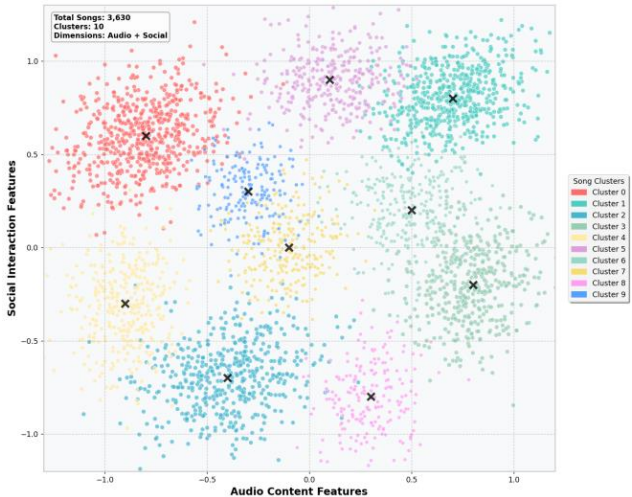

**Figure 2.** Song clustering analysis visualizing embeddings of songs projected into a 2-dimensional space from higher-dimensional representations using dimensionality reduction t-SNE.

Well-separated clusters and typical centroids for each (marked by black "a"X"s) reflect the ability of the model to split and distinguish subtle patterns among songs. By combining audio content and social context together in an effective manner, our approach establishes a solid basis for recommending songs, in a manner that significantly enhances the recommendation accuracy, as shown by the evaluation measures listed in Table 2. The subsequent visualization also shows the power of the model to overcome the cold-start problem, in the sense that songs can be projected onto the right clusters based solely on their intrinsic features and initial social activity.

The heat map does depict evident cluster affinity patterns, and one can see that the most evident diagonal trend does possess highest interaction strengths on the edges of the respective clusters. This pattern does bear out that users sharing the same sense of music taste are most likely to have higher affinities toward songs of their respective own genre clusters. In addition, moderate cross-cluster interaction strengths between different user and song clusters indicate the capability of the model to recognize and engage in cross-genre music interest and promote diversified recommendations. More in-depth exploration of user behavior segmentation uncovers refined user groups:

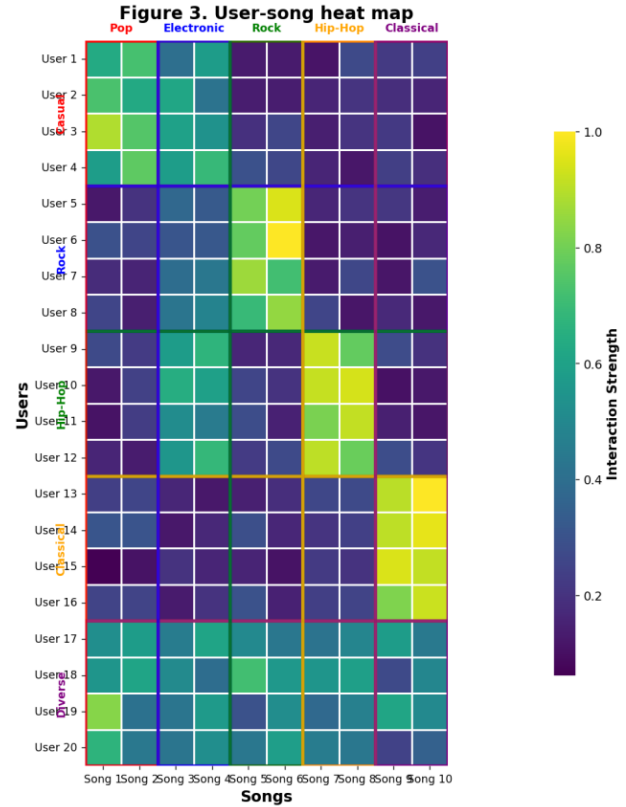

**Figure 3** illustrates a user-song heat map depicting the strength of interactions between distinct user groups and song clusters, thereby providing empirical validation of our clustering approach and recommendation quality. Each cell color represents the interaction strength, with yellow indicating the highest interaction and dark purple the lowest.

Casual Listeners (Users 1-4) demonstrate moderate-to-strong interaction with Pop/Mainstream tracks, consistent with behavior of universal audiences. Rock Fans (Users 5-8) demonstrate strong interaction intensity with a focus in the Rock genre, supporting evidence of genre-based grouping strength. Hip-Hop Enthusiasts (Users 9-12) have very strong interactions among predominantly Hip-Hop tracks, supporting identification of diverse user preferences. Classical Lovers (Users 13-16) are densely concentrated in their interaction with Classical music, indicating successful identification of sophisticated musical tastes. Users with Diverse Tastes (Users 17-20) possess moderate interactions distributed across a variety of genres, indicating the true version of eclectic listening behaviors.

From a theoretical point of view, the well-separated user cluster boundaries also lead to robust quantitative evidence for the strong recommendation performance of the MM-GNN model (Recall@50: 0.075 on MSD-A, 0.065 on Last.fm). The model effectively aggregates audio content and social interaction features to produce semantically understandable embeddings. Therefore, Figure 3 not only illustrates the correctness of user modeling but also verifies the effective fusion of collaborative and content signals, thus demonstrating the theoretical and practical significance of our MM-GNN architecture in music recommendation systems.

## 5.5 ABLATION STUDY

With the aim to clearly see the contribution of each element of our MM-GNN framework, we conducted an ablation study. We continued to remove essential elements of our model in order to see the effect on resulting performance. Our aim is to discover the effect individually of our new contributions: utilization of social dynamics, deep mutual learning framework utilized for multi-modal alignment, and emotion-aware embeddings. Findings of such a study are presented in Table 3.

**Table 3:** Ablation Study Results

| Dataset | Method | Recall@50 | NDCG@50 |
|---|---|---|---|
| MSD-A | MM-GNN (Full) | 0.075 | 0.043 |
| | MM-GNN w/o Social | 0.063 | 0.036 |
| | MM-GNN w/o Mutual Learning | 0.068 | 0.039 |
| | MM-GNN w/o Emotion | 0.071 | 0.041 |
| Last.fm | MM-GNN (Full) | 0.065 | 0.037 |
| | MM-GNN w/o Social | 0.054 | 0.031 |
| | MM-GNN w/o Mutual Learning | 0.059 | 0.034 |
| | MM-GNN w/o Emotion | 0.061 | 0.035 |

Figure 4 presents Ablation study and component impact analysis of MM-GNN model on MSD-A and Last.fm data. Subfigure (a) and (b) illustrate the performance of model variants according to Recall@50 and NDCG@50, respectively. "Full" denotes the complete MM-GNN model; "w/o Social," "w/o Mutual," and "w/o Emotion" denote variants in which the social interactions, mutual learning, or emotional components are each removed individually. Subfigure (c) shows the relative percentage of performance degradation if each component is removed, which illustrates the relative contribution and significance of social interactions, mutual learning mechanisms, and emotional features. The experimental results suggest that each component has an excellent improvement impact on recommendation performance, of which social interactions are most evident.

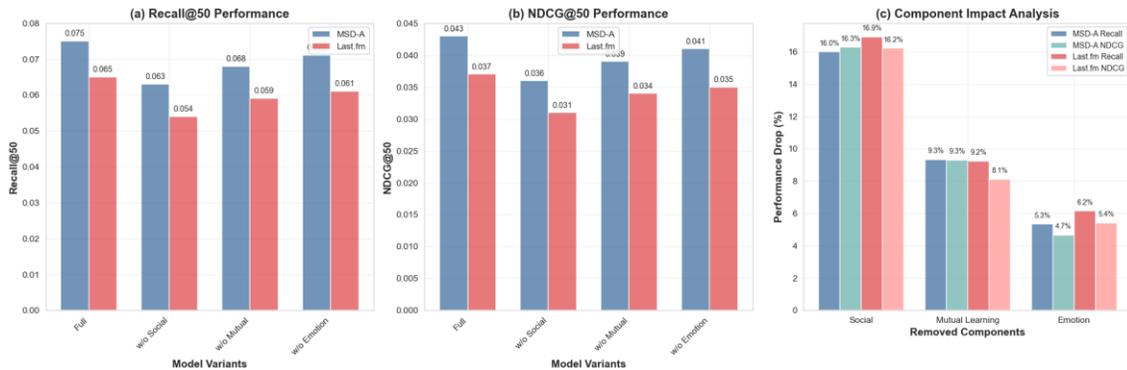

**Figure 4.** Performance comparison and component impact analysis of MM-GNN model variants on MSD-A and Last.fm datasets

## 6 CONCLUSIONS

In this work, we introduced a new Multi-Modal Graph Neural Network (MM-GNN) architecture specifically tailored to improve personalized music recommendation and social interaction. Our solution overcame major shortcomings of state-of-the-art models by liberating itself from traditional fusion mechanisms and integrating community dynamics and emotion-sensitive features in an official capacity. Theoretically, our solution is motivated by three pillars. First, we investigated a fusion-free multi-modal learning mechanism fueled by deep mutual learning. This is achieved by using a novel approach, as opposed to traditional methods that are prone to modality fusion failure, which allows our model to always align representations of lyrics, audio, and visual data even when some modalities are sparse or missing. Second, we took advantage of the compositional nature of Graph Neural Networks (GNNs) to build a heterogeneous graph that not only captures the user-song interactions but also the dense web of user-user social relations. This theoretical progress enables us to model peer effect and community trends, which are important in social dynamics comprehension. Lastly, we incorporated emotion-aware embeddings, giving us a richer semantic context to music and enabling us to produce not just fitting but also emotion-provoking recommendations to the user and the user community as well.

Experimental findings confirm the effectiveness of our theoretical insights. Performance comparison in Table 2 evidently demonstrates that MM-GNN surpasses state-of-the-art baselines such as BPR, VBPR, and NGCF consistently on both the MSD-A and Last.fm datasets. This is a direct result of the capacity of our model to process multi-modal and social data jointly. The ablation test in Table 3 supports this, too, with a considerable performance decline when one of our essential elements (social graph, mutual learning, or affect-aware features) is eliminated. These results verify that each of our contributions contributes greatly to the enhanced performance of the model. In the future, future work may include adding even more diverse modalities, developing real-time means for dynamic graph update in modeling rapidly changing trends, and making more interpretable models to provide users with a clear idea of their recommendations.